# Operating a Multi-Level Molecular Dimer Switch through Precise Tip-Molecule Control

Yueqing Shi[1,*], Weike Quan[1,2,*], Liya Bi[1,2], Zihao Wang[1,2], Kangkai Liang[1,2], Hao Zhou[1], Zhiyuan Yin[1], Wan-Lu Li[2,3] and Shaowei Li[1,2,†]

[1] Department of Chemistry and Biochemistry, University of California, San Diego, La Jolla, CA 92093-0309, USA

[2] Program in Materials Science and Engineering, University of California, San Diego, La Jolla, CA 92093-0418, USA

[3] Aiiso Yufeng Li Family Department of Chemical and Nano Engineering, University of California, San Diego, La Jolla, CA 92093-0448, USA

**Abstract**

Controlling structural transitions between molecular configurations is crucial for advancing functional molecular electronics. While reversible switching of bistable two-state molecules has been achieved, creating molecular systems that can be controllably switched between multiple configurations often requires complex synthetic methods, presenting a much greater challenge. In this study, we showcase a straightforward yet effective strategy to create and control transitions between multiple molecular structural states by forming a surface-bound molecular dimer. Using low-temperature scanning tunneling microscopy, we induce and characterize the structural transitions of a pyrrolidine dimer on a Cu(100) surface. The intermolecular interactions open new energy transfer channels, enabling excitation through pathways that were inaccessible in monomers. The occupation of different molecular states is highly sensitive to both the energy of the tunneling electrons and the interaction with the STM tip. By precisely adjusting the tip-molecule distance, we can select the most probable structural configuration based on sample bias, thereby achieving on-demand control of this molecular dimer switch. This work highlights an approach that leverages both intermolecular and molecule-cavity interactions to create and control an artificially fabricated molecular device.

\* These authors contribute equally to this work.

† Corresponding information: shaoweili@ucsd.edu

**Introduction**

Molecular electronics is a promising frontier for developing alternatives to traditional silicon-based technologies[1-5]. Precise control over transitions between molecular structural[6-10], electronic[11-13], or spin[14,15] states offers a range of strategies for realizing various molecular fundamental devices[16-19]. Among these approaches, exploiting molecular structural transitions, where molecules switch between two or more stable configurations, has shown potential as a key component in applications such as data storage and logic gates[20-24]. While reversible transitions between bistable two-state molecular systems — such as proton transfer[25,26], cis-trans transition[27-30], and ring-opening/closing[31-33] — occur naturally across various molecular systems, materials that can switch between multiple stable configurations are rare in nature. Designing and synthesizing the candidates of a multi-level molecular switch remains a significant challenge due to the complexity of creating molecules with the desired structures[34-36].

Scanning tunneling microscopy (STM) has become an indispensable tool for developing single-molecule functional devices because of its ability to resolve, characterize, and manipulate surface-adsorbed molecules with sub-angstrom precision. STM's high spatial resolution enables the detection of the changes in molecular states due to structural and electronic rearrangements[10,16,26,37-42]. Additionally, STM can deliver external stimuli, either by injecting energetic tunneling electrons or by applying external forces through the tip[43,44], to precisely excite the transition of the single molecule in the STM junction. More importantly, STM can manipulate surface adsorbates to create artificial structures[45-50], enabling the atomic-scale synthesis of molecular clusters with functionalities that individual molecules do not possess.

In this study, we demonstrate the controlled switching between multiple structural states of an artificially fabricated molecular dimer. Individual pyrrolidine molecules adsorbed on Cu(100) can be manipulated into dimers with a specific separation. We utilize STM to induce and control the transitions between various configurations of this pyrrolidine dimer. Our findings reveal that this pair of molecules can adopt six distinct configurational states, which can be switched from one to another when excited by tunneling electrons. Intermolecular interactions enable energy transfer between the two molecules, allowing excitation through pathways not accessible to monomers. By harnessing both intermolecular coupling and tip-molecule interactions, we were able to distinguish and control these molecular states through precise tuning of the tunneling electron energy and tip-molecule distance.

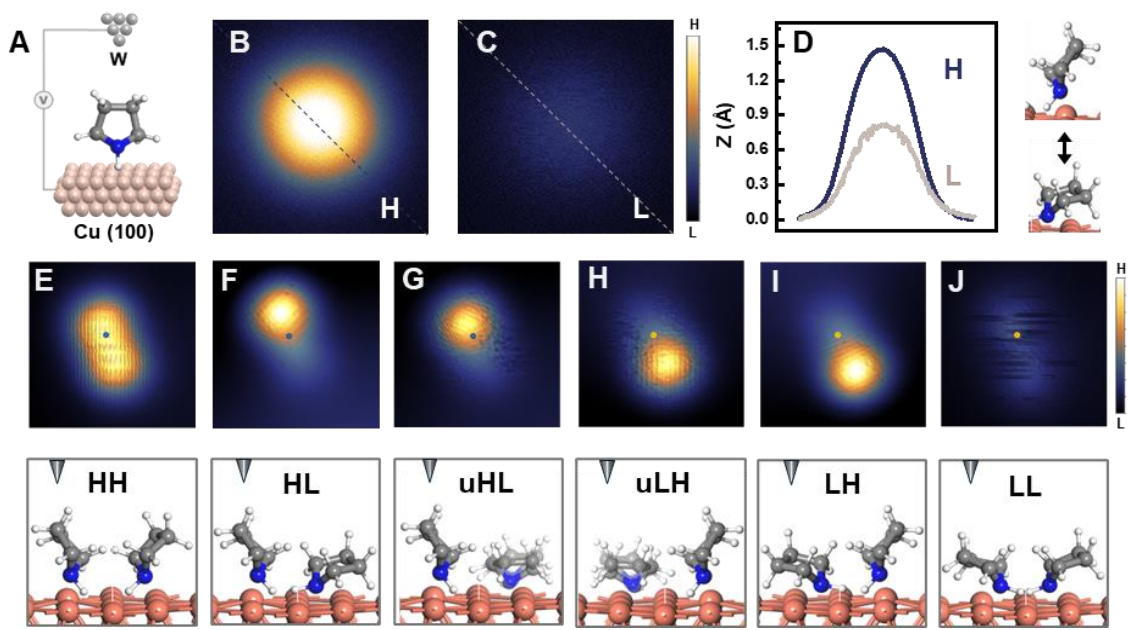

Figure 1. Multiple structural states of pyrrolidine monomer and dimer on the Cu(100) surface. (A) Schematic diagram of a pyrrolidine monomer adsorbed on Cu(100) probed with STM. (B)-(C) Topographic images revealing the High (B) and Low (C) states of pyrrolidine monomer. Image scale: 1.3 nm × 1.3 nm. (D) Line-cut profiles of the two states along the dashed line in (B) and (C). The monomer can reversibly switch between these two states when excited by tunneling electrons. (E)-(J) Topographic images revealing the six structural states of a pyrrolidine dimer. Image scale: 1.4 nm × 1.4 nm. (E) High-High (HH) state, (F) Stable High-Low (HL) state, (G) Unstable High-Low (uHL) state, (H) Unstable Low-High (uLH) state, (I) Stable Low-High (LH) state, (J) Low-Low (LL) state. The circle indicates where the tip is positioned to record the current-time trace. The tunneling gap was set at 50 mV sample bias, 20 pA tunneling current, with the feedback on for Figures (B)-(C), and with the feedback off for Figures (E)-(J).

**Results and Discussion**

The building block of our multi-level molecular switch is the single pyrrolidine molecules, characterized by its simple five-membered saturated ring consisting of one nitrogen and four carbon atoms (**Fig. 1A**). The pyrrolidine monomer is known to have two distinct structural conformations when adsorbed on Cu(100), standing and bent toward the surface, namely the high (H) state and the low (L) state, which are associated with varying apparent heights as depicted in **Fig. 1B-D**. Existing reports indicate that the excitation of specific vibrational modes by inelastic tunneling electrons enables the pyrrolidine monomer to reversibly transition between these two structures[51,52], serving as a simple example of a bi-level molecular switch[53]. When adsorbed on the four-fold symmetric Cu(100) surface, pyrrolidine molecules exhibit rapid reversible rotation among four equivalent orientations (**Fig. S1**), resulting in a circular appearance in both high and low states.

Pyrrolidine dimers can be formed naturally at a relatively high molecular coverage or through artificial manipulation with the STM tip following established techniques[45-49]. On the Cu(100) surface, pyrrolidines can be arranged into pairs with varying intermolecular distances. When two molecules are relatively far apart, each molecule switches independently, showing no significant correlation. However, when two molecules become too closely packed, they form a single, stable configuration that is unable to change further. In the specific type of dimer with intermolecular separation of ~0.4 nm investigated in this study, we observed six distinct states in both topographic images and tunneling current traces — exceeding the four states expected from the simple combinations of the two states of each monomer. Besides both molecules in the high state (HH), and both in the low state (LL), we observe two pairs, four possible states corresponding to one molecule in the high state and the other in the low state. One pair shows relatively scraping signals in the topographic image (**Fig. 1G and 1H**), making it appear less stable. They are referred to as the Unstable High-Low (uHL)/Low-High state (uLH) states, in contrast to their more stable counterparts termed High-Low (HL)/Low-High (LH) states (**Fig. 1F and 1I**). The first capital letter in the abbreviation denotes the state of the molecule closer to the STM tip. The discrepancy between the stable and unstable configurations likely results from the steric repulsion between molecules in proximity, which breaks the energy degeneracy of the four in-plane geometries of the molecule in the L configuration (see **Supporting Information Section 3.1** for more detailed discussion). The unstable states are much shorter-lived, typically appearing as intermediate steps during transitions between stable configurations. Thus, we primarily utilize the four stable configurations as the bases to fabricate a multi-level molecular switch.

To create a functional molecular switch, it is essential to reliably switch the molecule into a desired state on demand. At the experimental temperature of 5 K, spontaneous structural transitions do not occur for either monomers or dimers but can be driven by inelastic tunneling electrons. For the monomers, the resulting configuration change can be tracked by monitoring changes between two tunneling current levels, with the feedback turned off and the STM tip positioned over the center of the pyrrolidine molecule. The occupation ratio of a state, defined as the average percentage of time a molecule remains in a specific configuration, varies with the sample bias, as shown in **Fig. 2A**. When the bias is set below 20 mV, the tunneling electrons cannot trigger any structural transition. Between 30-40 mV, the molecule can be switched to and stays at the H state. Beyond this voltage region, the occupation ratio of the L state increases, reaching its peak at 60 mV. As the voltage increases further, a resurgence of the H state occurs, becoming the dominant state once again, as illustrated in **Fig. 2B**.

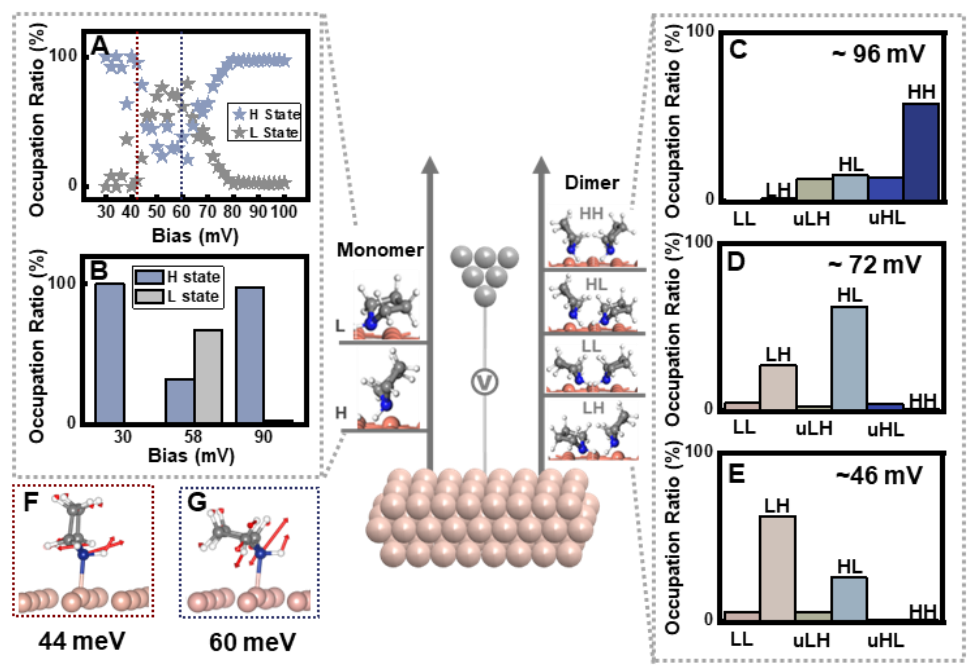

Figure 2. Occupation ratio of different molecular configuration as a function of sample bias. (A) Voltage-dependent occupation ratio the L and H states of a pyrrolidine monomer. The tunneling gap was established at a sample bias of 50 mV and a current of 20 pA, with the tip retracted 0.02 nm away from the molecule. (B) Statistic histogram of the occupation ratio of monomer L and H states at 32/58/90 mV. (C)-(E) Occupation ratio of different dimer configurations (C)96 mV, (D)72 mV, and (E)46 mV. The tunneling gap was set at 50 mV sample bias, 20 pA tunneling current. (F)-(G) DFT simulations of the two vibrational modes (F)of the H state molecule related to the first H-to-L transition threshold at 44 meV and (G) of the L state molecule related to the first L-to-H transition threshold at 60 meV, respectively.

The bias dependence of the H and L state occupation ratio is closely linked to the dominant transition pathways excited by inelastic electron tunneling. To identify the nature of these pathways, we analyze the transition rate as a function of sample bias from current-time traces, which represent the probability that a tunneling electron at a given bias could trigger the transition. As shown in **Fig. S2**, the transition rates display several distinct jumps at specific bias values, corresponding to the excitation of different vibrational modes that open new pathways for H-to-L or L-to-H transitions. By fitting the data using established formulas[54,55] (detailed in **Supporting Information Sections 1.2 and 3.2**), we determine the H-to-L transition thresholds to be approximately 44 meV and 78 meV, and the L-to-H thresholds to be around 60 meV and 75 meV. The competition between these transition pathways determines the overall H and L occupation ratio of the molecule at different biases. When the bias is below 20 mV, the electron energy is insufficient to overcome the energy barrier between the L and H states, so the molecule remains in its initial configuration. In the 30 to 40 mV range, electrons occasionally trigger the L-to-H transition, locking the molecule to its more energetically favorable H state. When the bias exceeds 44 mV, the excitation of a vibrational mode in the H state molecule (**Fig. 2F**) activates an H-to-L transition channel, increasing the occupation ratio of the L state. As the voltage approaches the L-to-H transition threshold of 60 mV, associated with a vibrational mode of the L state molecule (**Fig. 2G**), a new L-to-H transition is triggered, leading to a decrease in the L state occupation ratio. Consequently, the molecular switch can be controlled by fine-tuning the energy of the tunneling electrons. For instance, the molecular configuration can be switched by applying a relatively high bias and then locked into that state by lowering the bias to below 20 mV. Therefore, identifying the specific bias at which the molecule predominantly occupies a desired configuration—such that it spends the majority of time in that state—is crucial for enabling consistent and reversible molecular switching.

To conveniently operate the dimer switch, it is important that transitions between all molecular configurations can be excited and monitored by tunneling electron injection at the same location. In the symmetric dimer we studied, we positioned the STM tip near one of the molecules, where we can identify all six distinct structural states by their different levels in the tunneling current. The tip's position is indicated by circles in **Fig. 1E-J**. Notably, the transitions between these six states exhibit directionality depending on the position of the tip. Direct transitions between certain states, such as between the HH and LL states or the HL/uHL and LH/uLH states, were not observed at voltages below 70 mV. (**Fig. S3**). These "forbidden" transitions all require simultaneous changes in both molecules, likely involving higher energy barriers or multiple electron excitations, making them less favorable.

Similar to the monomers, the occupation ratio of different dimer states varies depending on the energy of tunneling electrons as illustrated in **Fig. S4**. At bias between 30 and 40 mV, the dimer preferentially adopts the HH configuration, aligning with the behavior of the monomer where the H configuration is the more energetically preferable. As voltage increases, the molecules start to be excited into to L state, leading to the dominance of HL, LH, and LL states within the 40-75 mV range. The uHL and uLH states were only observed as the short intermediate step for the transition from HH to HL or LH states. The occupation ratios as a function of sample bias for both the HL and LH states exhibit a double-peak feature. In both states, two occupation ratio maximums occur at approximately 45 meV and 70 meV. Although sharing the same peak energy, the relative occupation ratio between these two states is different. At approximately 45 mV, the dimer predominantly occupies the LH state, with an occupation ratio of about 63% (**Fig. 2E**), whereas at around 72 mV, the HL state is favored with an occupation ratio of approximately 62% (**Fig. 2D**). The LL state has a maximum occupation ratio at around 60 mV, coinciding with the monomer case where the L-to-H transition becomes preferable. As the voltage increases further, the HH state becomes dominant once again (**Fig. 2C**). Note that even though the occupation ratio of LL maximizes at around 60 mV, it is still merely comparable to HL, and LH states (**Fig. S4**). However, the occupation ratio of the other three stable states, HH, HL, and LH, can all be easily distinguished at different biases.

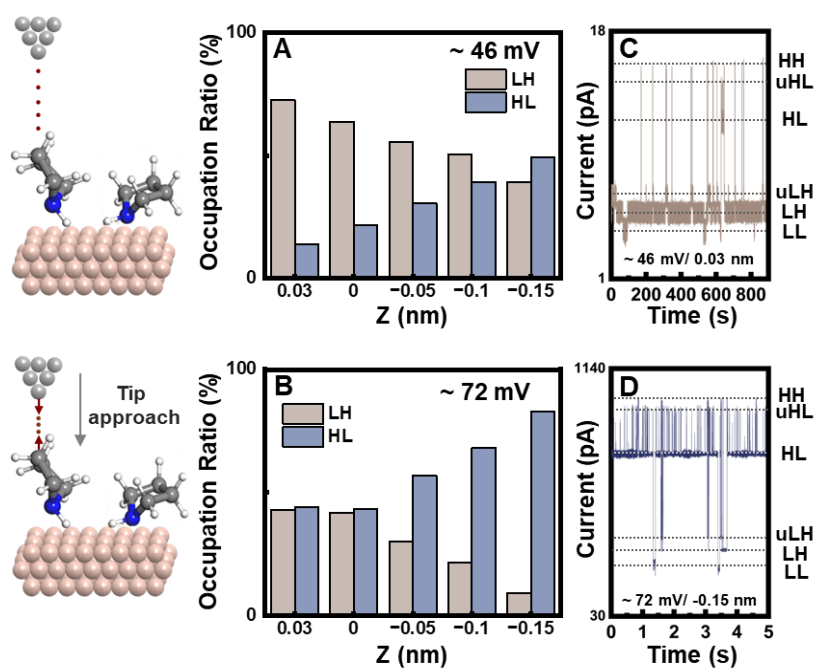

Figure 3. Alternation to the HL and HL state occupation by tip-molecule interaction. (A)-(B) Occupation ratio of the dimer HL and LH states at (A) 46 mV and (B) 72 mV at different tip molecule separation. The reference tunneling gap (Δz=0 nm) is set at 50 mV sample bias, 20 pA tunneling current. (C)-(D) The current traces taken at (C) Δz=0.03 nm and bias=46 mV, (D) Δz=-0.15 nm and bias=72 mV. Positive Δz refers to retracting the tip away from the molecule, negative Δz means approaching the tip toward the molecule.

The HL and LH states are inherently symmetric, which is broken by the presence of the STM tip. **Fig. 3** illustrates the effect of the tip-molecule distance on the occupation ratio of the HL and LH states of the dimer. When the tip is farther from the molecules, the dimer strongly favors the LH state at ~46 mV, with an occupation ratio of up to 75% (**Fig. 3A and 3C**). As the tip moves closer to the sample, the proportion of the LH state steadily decreases at both 46 mV and 72 mV, with the HL state becoming increasingly dominant. At minimized tip-sample distance, the HL state predominates at 72 mV, reaching an 83% occupation ratio (**Fig. 3B and 3D**). Thus, by modulating the tip-molecule interaction, HL and LH states can be easily differentiated by their occupation ratio at these two biases.

To understand the mechanism behind the operation of this multi-level dimer switch, three key factors must be considered: simultaneous excitation of both molecules in the dimer, tip-molecule interaction, and intermolecular couplings. If we model the dimer as two independent monomers excited simultaneously by tunneling electrons, the occupation ratio of different dimer states can be represented as the product of the probabilities of the H and L states of each monomer at various bias voltages, detailed discussion is shown in the **Supporting Information Section 3.3**. Based on the monomer data in Figure 2A, the occupation probability of the L state can be approximated by a Gaussian function, $P(V)$, centered around 58 mV, as shown in **Fig. S5**. Consequently, the probability of the H state is $1-P$. This leads to the LL state being expressed as $P^2$, and the HL and LH states as $P(1-P)$. The tip-molecule and intermolecular interaction are not considered in this model. According to this simplification, the bias dependence of the HL and LH occupation ratio is expected to exhibit a split double-peak feature, aligning with experimental observations (**Fig. S4**), while the LL state retains a single peak centered around 58 mV. This naturally provides a mechanism to differentiate HL/LH and LL at different applied bias voltages.

To further tell between HL and LH, it is important to consider how the STM tip interacts differently with the two molecules. Previous studies have shown that the tip exerts an attractive van der Waals force on the molecule directly beneath it, lowering the energy of the H configuration for that molecule[53]. Therefore, as the tip approaches the molecule, the molecule closer to the tip prefers the H configuration, increasing the probability of the HL state over the LH at both 46 mV and 72 mV, as shown in **Fig. 3A and 3B**. This allows this dimer switch to be easily switched to the HL state at 72 mV at a small tip-molecule separation.

However, there is an interesting discrepancy between the experimental observation described in **Fig. 3A** and the simplified model described in **Fig. S5**: the LH state is much more favorable than the HL state at 46 mV even with a considerably large tip-molecule separation. If the HL and LH are solely differentiated by the tip-molecule van der Waals attraction, they should become less distinguishable when the tip is far away. Yet, as shown in **Fig. 3A**, when the tip is retracted from the molecule, the LH state achieves an occupation ratio of up to 75%, while the LH state occupation ratio is below 10%.

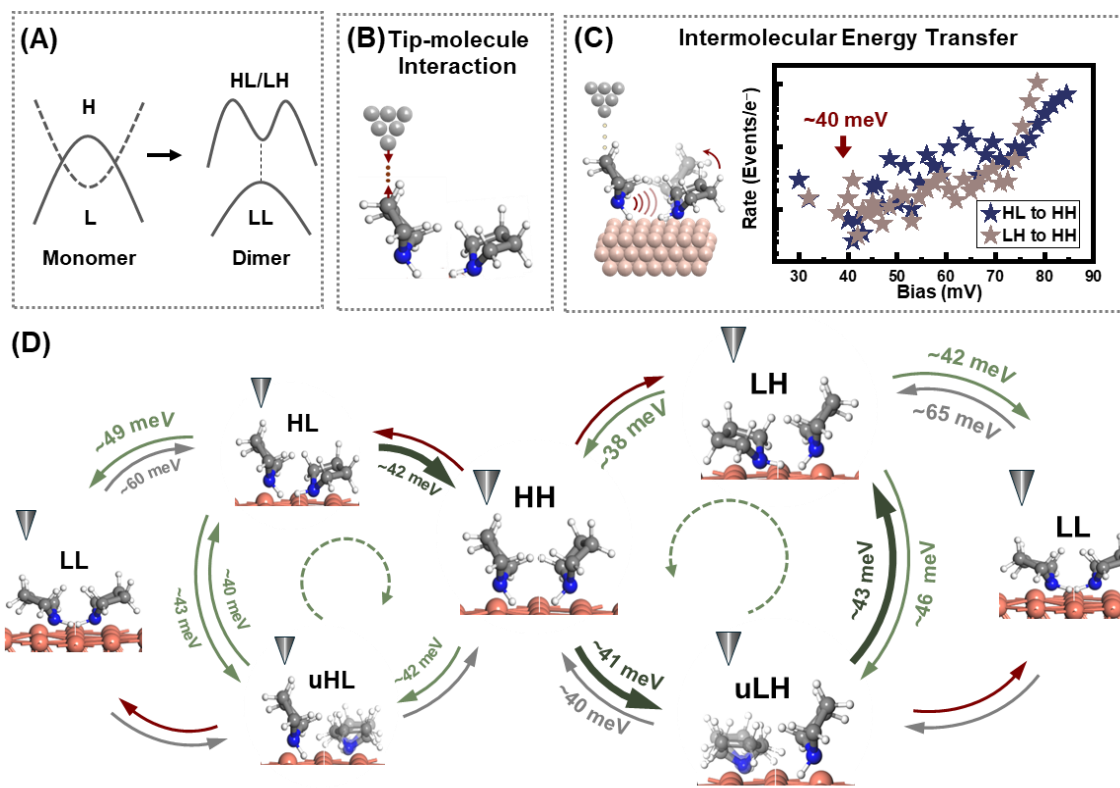

Figure 4. Schematic of reaction pathways of the dimer switching that can be activated by the inelastic tunneling electrons. (A) A simplified model estimating the dimer occupation ratio based on the occupation ratio of H and L of a monomer. (B) Schematic diagram indicating how the interaction between the STM tip and the molecule underneath it is influencing the preference of difference dimmer configuration. (C) The HL/LH to HH transition rate as a function of bias. The lowest energy threshold is measured around 40 meV. (D) Schematic diagram of the directional transition between different dimer states and the corresponding energy thresholds extracted from the experimental measurement. The green arrows indicate that this transition can readily occur within the 40-60 mV range. Gray arrows indicate that the transition can occur but with a much lower chance. Red arrows indicate that the transition is rarely observed. The thicker green arrows indicate the most preferrable transition path when bias is set around 46 meV, forming a directional transition pathway: HL → HH → uLH/uHL → LH.

To understand this discrepancy, we analyzed the directional transition between different dimer configurations at varying bias voltages (**Fig. S6**). The identified transition thresholds between different states are shown in **Fig. 4D**. When the bias is set between 40~50 mV, the dimer favors a unidirectional cyclic transition: HH-LH/HL-HH. Detailed discussion regarding how this directional transition is determined is included in the **Supporting Information Sections 3.4 and 3.5**. Interestingly, it appears that the HL-HH-LH transition is much more favorable than LH-HH-HL transition, eventually leading to the dominant occupation of LH.

The directional transition from HL to LH is facilitated by interactions between the two molecules, which open new energy transfer channels and create distinct excitation pathways that are absent in the transitions of monomers. In monomers, the first transition pathway from L state to H state is triggered by the excitation of a vibrational mode in the L state molecule at ~60 mV. However, when molecules form a dimer, the transition from HL/LH states to HH state—also involving the transition of one molecule from L state to H state—shows a threshold energy at ~40 meV. This lower energy threshold, around 40 meV, differs from the thresholds observed for the monomer's L-to-H transition (**Fig. S2A**) and the dimer's LL-to-HL/LH transition (**Fig. S5M and S5N**). However, it aligns with the vibrational mode of the H-state molecule, a mode not accessible in the monomer's L-to-H transition.

This new lower-energy transition threshold suggests the existence of an energy transfer channel where the vibrationally excited H state molecule in the HL or LH configuration transfers energy to the neighboring L state molecule, prompting its transition from L to H, likely resulting from the anharmonic coupling between the vibrational states of two molecules. As shown in **Fig. 4C**, the transition probability from HL state to HH state is significantly higher than that from LH state between 40-70 mV. This discrepancy is likely related to the excitation probability of the H state vibration, which depends on the position of the STM tip. When the tip is positioned above the H state molecule of the dimer in the HL state, the cross-section for exciting the H state vibration is much greater than in the LH case, where the tip is positioned over the L state molecule. Additionally, since the H-to-L transition can be mediated by the same vibration, the HH-to-LH transition is also facilitated, during which the H-state molecule under the tip switches to the L state. These sequential preferential excitations create a directional LH-to-HL pathway when the bias is set to ~46 mV, even at a considerable tip-molecule separation, due to the continuous excitation of the vibration of the H state molecule.

**Conclusion**

In conclusion, our study presents an efficient and straightforward approach to on-demand control of a multi-state molecular switch utilizing a pyrrolidine dimer. The formation of a dimer introduces additional structural states without the involvement of complicated synthetic approaches. By positioning the STM at a location closer to one of the molecules in the dimer, we can not only monitor all different molecular structures but also break the energy degeneracy of some of these states via the tip-molecule interaction, allowing the differentiation of molecular configurations which are otherwise indistinguishable in energy. Intermolecular interactions introduce a new energy transfer channel, allowing directional transfer between dimer configurations. By precisely adjusting the applied voltage and the tip-molecule distance, we have successfully presented a material candidate for a controllable multi-state molecular switch. This research elucidates the possibility of utilizing the fundamental molecule-environment interaction to manipulate the transition between molecular conformations, paving the way for designing more complex and efficient molecular functional devices with artificial molecular clusters.

## Acknowledgment

This work was supported by the United States National Science Foundation (NSF) under Grant No. CHE-2303936 (to Shaowei Li). The authors acknowledge the use of facilities and instrumentation supported by NSF through the UC San Diego Materials Research Science and Engineering Center (UCSD MRSEC) with Grant No. DMR-2011924.